\documentclass[prl,twocolumn,showpacs]{revtex4}
\usepackage{graphicx}
\begin{document}
\title{Susceptibility of the transverse field Ising model on the square lattice}
\author{A. Kashuba}
\affiliation{Bogolyubov Institute for Theoretical Physics, 14-b Metrolohichna, Kiev 03680 Ukraine}

\begin{abstract} Susceptibility of the transverse field Ising model on the square lattice is calculated numerically in the paramagnetic phase in a wide range of temperatures and transverse fields. An expression with one constant $\pi$, that determines both the critical exponent $\gamma$ and the critical transverse field, compellingly represents the data asymptotically near the quantum critical point, except for a narrow classical region close to the phase transition line, and shows two crossovers as temperature varies. \end{abstract}


\maketitle

Quantum field theory having been derived from an explicitly Lorentz invariant model encounters divergences when imposing quantum mechanics. Alternatively, quantum models defined on lattices explicitly obey the quantum mechanics but the Lorentz invariance, the conformal symmetry as well as other symmetries emerge, if at all, only in the long wavelength limit \cite{polyakov}. This continuum limit can be found either in special massless phases, such as quantum antiferromagnets, or in the vicinity of a phase transition known as a quantum critical point at zero temperature. Novel emergent phenomena near quantum critical points is of great interest in condensed matter physics \cite{sachdev}. Phase diagrams near quantum critical points show abundant characteristic crossovers \cite{sach98} like a pseudogap phenomenon in the high-temperature superconductors. Symmetry brings in simplicity. First demonstrated by Maldacena \cite{maldacena}, quantitative description of quantum lattice models governed by a symmetry may correspond to a solution of the classical Einstein equation for gravitational fields in special settings representing the same symmetry. It is plausible that such a solution in the case of the quantum critical point would be given in terms of a simple, 'school-curriculum' function. Provided this simplicity, underpinned by unknown symmetry, precise numerical simulations may be sufficient to guess the answer for some physical questions. This paper, based on numerical data, reveals that in the case of the transverse field Ising model on the square lattice the magnetic susceptibility can be represented as a simple expression in a wide range of temperatures and transverse fields around the quantum critical point. In particular, both the position of the quantum critical point and the critical exponent of the magnetic susceptibility is specified. There are two crossovers at $T^*_1=1/\pi$ and $T^*_2=\pi$ in the expression for magnetic susceptibility.

The spin-half transverse field Ising model \cite{pfeuty} on the square lattice is anisotropic in the spin space with two distinct axis, longitudinal $x$ and transverse $z$. The Hamiltonian includes the exchange interaction between nearest neighbors $\langle \mathbf{x} \mathbf{y} \rangle$, the Zeeman energy in the transverse field $H$ and the external longitudinal magnetic field $h$:
\begin{equation}\label{Ham}
\hat{H}= -J\sum_{\langle \mathbf{x} \mathbf{y} \rangle}\ \sigma^x_{\mathbf{x}}\ \sigma^x_{\mathbf{y}} -H\sum_{\mathbf{x}} \sigma^z_{\mathbf{x}} -h\sum_{\mathbf{x}} \sigma^x_{\mathbf{x}}\ ,
\end{equation}
where spin-half $\vec{\sigma}_{\mathbf{x}}$ resides on sites $\mathbf{x}$ of the square lattice. We set the exchange coupling to unity, $J=1$, leaving a single parameter, the strength of the transverse field $H$, in the Hamiltonian eq.(\ref{Ham}). Without an external field the Hamiltonian does commute with the parity operator: $\hat{P}=\prod_{\mathbf{x}} \sigma^z_{\mathbf{x}}$. Accordingly, all eigenstates of this model are divided into even and odd states of equal number. Available experimental realizations of the transverse field Ising model either include long range interactions \cite{bra} or are one-dimensional magnetic chains \cite{coldea}. 

\begin{figure}
\includegraphics[width=8cm]{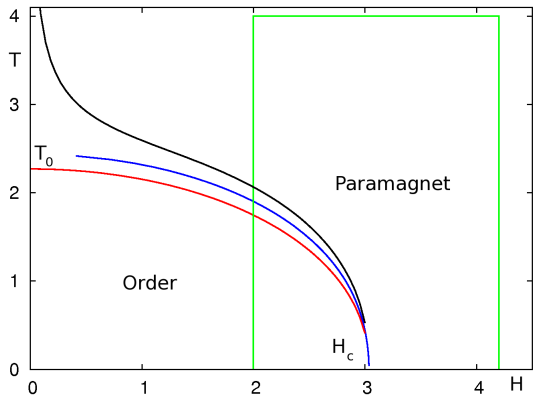}
\caption{\label{fig:diag} Phase diagram $(H,T)$ of the transverse field Ising model. Inside the green box the susceptibility data have been collected. Red, lower, line is the Ising phase transition $T_c(H)$ ending in the quantum critical point $H_c$. Blue, middle, line is a spurious quantum transition $T_Q(H)$ [singularity in eq.(\ref{QCfit})] ending before reaching $H=0$ axis. Black, upper, line is a concept of the quantum-to-classical crossover $T^*(H)$.}
\end{figure}

Phase diagram $(H,T)$ of the transverse field Ising model on the square lattice is shown schematically on Fig.\ref{fig:diag}. There are two phases, a paramagnetic phase at high temperatures and high transverse fields and a magnetically ordered along the longitudinal direction phase at low temperatures and low transverse fields. There is always a non-vanishing magnetization in the transverse direction that grows with $H$. A line of phase transitions $(H,T_c(H))$, not found in this paper, separates these two phases. On the right, it ends in a quantum critical point, $(H_c,0)$, and, on the left, in a classical critical point, $(0,T_c(0))$, the phase transition of the classical two-dimensional Ising model at $T_0=2/\textrm{asinh}(1)=2.269185$. At the approaches to the phase transition line the longitudinal susceptibility diverges according to a power law: 
\begin{equation}\label{chiCl}
\chi(H,T)=C^+_{cl}(H)\left( \frac{T}{T_c(H)}-1 \right)^{-7/4}\ ,
\end{equation}
corresponding to the universality class of the classical two-dimensional Ising model \cite{bmcw}. On the line $T=0$, the quantum two-dimensional transverse field Ising model is equivalent to the classical three-dimensional Ising model \cite{suzuki,fs78}. At the approaches to the quantum critical point along the line $T=0$ the longitudinal susceptibility diverges as
\begin{equation}\label{chiQ}
\chi(H)=C^+_{Q}\left( \frac{H}{H_c}-1 \right)^{-\gamma}\ ,
\end{equation}
where $\gamma$, found to lie in the interval $\gamma=1.23 ... 1.24$ in many studies both quantum and classical \cite{bc97,gzj98,hasenbusch,hamer}, is a critical exponent of the susceptibility in the classical three-dimensional Ising model.

We calculate numerically the longitudinal susceptibility using Nickel's linked cluster expansion method \cite{gs} called as a graph expansion below. It applies to all quantum lattice models with the Hamiltonian being a uniform sum over lattice edges:
\begin{equation}\label{bond}
\hat{H}=\sum_{\mathbf{e}} \hat{H}_{\mathbf{e}}=\sum_{\langle \mathbf{x}\mathbf{y} \rangle} \left[ -\frac{H}{4}\left(\sigma^z_{\mathbf{x}}+\sigma^z_{\mathbf{y}} \right)-\sigma^x_{\mathbf{x}}\ \sigma^x_{\mathbf{y}} \right] \ ,
\end{equation}
and to the transverse field Ising model in particular. The transverse field Zeeman term as well as the longitudinal magnetization operator:
\begin{equation}\label{longmag}
\hat{M}= \sum_{\langle \mathbf{x}\mathbf{y} \rangle} \frac{1}{4} \left(\sigma^x_{\mathbf{x}} +\sigma^x_{\mathbf{y}} \right)\ ,
\end{equation}
for each site can be split into four equal parts assigned to the four incident edges. Our algorithm would not require it but in derivation we assume the validity of the perturbation theory, for the transverse field Ising model in the limit $H\to\infty$. The transverse field Zeeman term is the unperturbed Hamiltonian whereas the Ising term, residing on edges, is the perturbation. Usually, perturbation terms are arranged by the power. Instead, we assign a footprint i.e. a set of connected edges involved in the given perturbation process. Different footprints specify classes of perturbation terms. In general, some perturbation terms are equal while having different footprints. Such footprints are different embedding of the same graph into the lattice. Truly distinct classes are represented by graphs embeddable into the lattice. The result of summing up the perturbative series on a graph is alternatively found using the matrix quantum mechanics. The Hamiltonian $\hat{H}_g$ and a longitudinal magnetization $\hat{M}_g$ of a graph $g$ are uniquely defined as restriction of the lattice Hamiltonian eq.(\ref{bond}) and the longitudinal magnetization eq.(\ref{longmag}) onto one embedding of this graph into the lattice. Both $\hat{H}_g$ and $\hat{M}_g$ do not depend on the particular embedding. The operator $\hat{M}_g$ connects even and odd states. The even and odd blocks of $\hat{H}_g$ as well as $\hat{M}_g$ have all equal dimension.

We arrive at the following algorithm. We enumerate all graphs embeddable into the square lattice. There are in total $1,1,2,4,6,14,28,68,156,399,1012$ graphs embeddable into the square lattice with the number of edges $1,2,3,4,5,6,7,8,$ $9,10,11$ correspondingly \cite{oeis}. For each graph $g$ the graph Hamiltonian $\hat{H}_g=\sum_i |i\rangle E_g(i)\langle i|$ is diagonalized numerically whereas the graph magnetization $\langle i|\hat{M}_g| j \rangle$ is rotated into the new basis. The Gibbs thermodynamic average of the graph susceptibility reads:
\begin{equation}
\chi_g =\sum_{i,j} \frac{\langle i|\hat{M}_g| j \rangle \langle j|\hat{M}_g| i \rangle}{E_g(j)-E_g(i)} e^{\displaystyle -E_g(i)/T}/ \sum_i e^{\displaystyle -E_g(i)/T}
\end{equation}
The Gibbs thermodynamic average of the magnetic susceptibility on the lattice per one site in the thermodynamic limit is a sum of the graph susceptibilities:
\begin{equation}\label{graph}
\chi =\sum_g Z(g)\left[ \chi_g - \sum_{f\in g} Z_g(f) \chi_f \right] \ ,
\end{equation}
where $Z(g)$ is the number of different embeddings of the graph $g$ into the lattice. $f$ is the sub-graph of the graph $g$, hence, also embeddable into the lattice. $Z_g(f)$ is the number of different embeddings of the sub-graph $f$ into the graph $g$ considered on the lattice.

The graph expansion alters in the structure of the Feynman diagrams method where one sums up 'geometry' in terms of propagators and interaction points first and gives the result in powers of the coupling. In the graph expansion one sums up the interaction in all powers of the coupling first and gives the result in geometrical patterns, footprints. Also, the graph expansion counts largely the same processes as the Lanczos method on the regular $6\times 6$ cluster albeit on many thousands of graphs bended and turned in many thousands ways with observables being given in the thermodynamic limit.

\begin{figure}
\includegraphics[width=10cm]{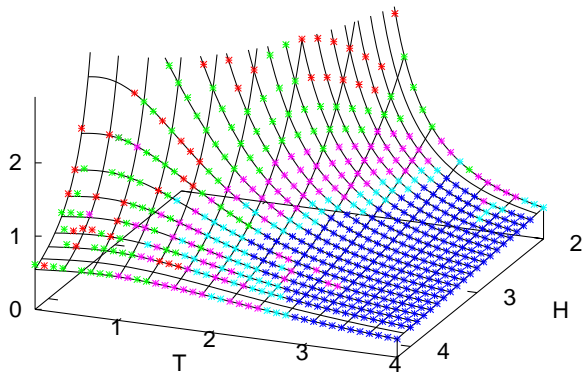}
\caption{\label{fig:susc} Longitudinal susceptibility sorted out into five quality grade baskets, dark blue 281 points, blue 73 points, pink 100 points, green 102 points and red 46 points vs $(H,T)$. The eqs.(\ref{chi},\ref{QCfit},\ref{regular}) is shown as the black surface.}
\end{figure}

Graphs are classified according to the number of edges, referred to here as a weight $w$. The sum eq.(\ref{graph}) restricted all graphs $g$ with the weight $w$ defines a partial susceptibility $\chi_w$. For each point $(H,T)$ on the phase diagram the susceptibility is given by a sum of partial susceptibilities, $\chi(H,T)=\sum_w \chi_w(H,T)$. A sequence of numbers $\chi_w(H,T)$ for $w=1 ... 11$ is calculated numerically. In the paramagnetic phase a typical sequence $|\chi_w|$ seems to be convergent. In the magnetically ordered phase a typical sequence $|\chi_w|$ seems to be divergent. At $H<H_c$ the sign of $\chi_w$ is positive whereas at $H>H_c$ the sign of $\chi_w$ shows typically an irregular pattern. To extrapolate to $w=\infty$ while allowing for a one change in the pattern, the available sequence is split in the middle at some weight $w^*$. Then, $\chi_w$ for $w<w^*$ is summed up whereas $\chi_w$ for $w\geq w^*$ is augmented by a variable $z$ into a polynomial:
\begin{equation}
\chi(H,T)=\sum_{w=1}^{w^*-1}\chi_w(H,T)+\sum_{w=w^*}^{w_{max}} \chi_w(H,T)\ z^{w-w^*}
\end{equation}
where $w_{max}=11$. This polynomial is extrapolated using the Pade approximation. Selecting all middle weights in the interval $2\leq w^*\leq 9$ and all degrees of the polynomial in the nominator of the Pade approximant while sending $z$ to one gives us forty or so different extrapolations $\chi_i$ for one point $\chi(H,T)$. Their distribution $\rho(\chi)$ has a maximum corresponding, probably, to the correct extrapolation $\chi(H,T)$, and a tail of those $\chi_i$ that are gone astray when $z=1$ comes close to a pole. We decimate the extrapolations most distant from the average $\langle \chi_i \rangle$, one by one, and stop when four extrapolations remain. Their average gives us $\chi(H,T)$. For a measure of quality of thus calculated datum the above procedure is repeated twice for $w_{max}= 10$ and $w_{max}= 11$. The absolute difference between the two to the datum value ratio defines a quality of the datum, $\delta(H,T)= |\Delta\chi|/\chi$.

Fig.\ref{fig:susc} shows the longitudinal susceptibility of the transverse field Ising model on the square lattice calculated numerically in the range $0.1\leq T\leq 4$ and $2\leq H\leq 4.2$ with a step $\Delta H=\Delta T=0.1$. The data with a poor quality $\delta(H,T)>3.2\%$ are not shown. $\chi_0=1/(2H_c)=0.165$ sets an atomic scale for the susceptibility. We sort out our data, six hundred points in total, into five quality baskets: $\delta \leq 0.05\% \leq \delta \leq 0.1\% \leq \delta \leq 0.4\% \leq \delta \leq 1.6\% \leq \delta \leq 3.2\%$, shown in dark blue, blue, pink, green and red colors correspondingly on the Fig.2. The closer one approaches the Ising phase transition line and especially the line $T=0$ the worse is the quality of the data. The graph expansion, relying on discreet energy spectra of small graphs, is expected to become problematic in the limit $T\to 0$.

In the paramagnetic phase the longitudinal susceptibility is given by the Kubo equation:
\begin{equation}
\chi(H,T)=\sum_{\mathbf{r}} \int_0^{\infty} \langle \left[\sigma^x(0,\mathbf{0}), \sigma^x(t,\mathbf{r}) \right]_-\rangle dt =\int_{0}^{\infty} \frac{d\chi}{d\xi} d\xi
\end{equation}
where in the spirit of the renormalization group \cite{polyakov} at large distance $|\mathbf{r}|\gg 1$ the sum proceeds in a scale-wise manner, with the scale $\xi=\log|\mathbf{r}|$. In classical statistical physics the susceptibility density $d\chi/d\xi$ is determined by a running renormalization group energy, Hamiltonian, $H_{RG}(\xi)$. As $\xi$ grows it approaches the fixed-point Hamiltonian $H_{FP}$ and, near a critical point, stays in the fixed-point, $H_{RG}(\xi)=H_{FP}$, for a long interval of $\xi$. Here a critical part of the susceptibility $\chi_{C}$ develops. At the initial transient scales as well as at the exit from the fixed-point a regular part of the susceptibility $\chi_{reg}$ develops. Analogously, we write for the quantum model:
\begin{equation}\label{chi}
\chi(H,T)=\chi_{QC}(H,T)+\chi_{reg}(H,T)
\end{equation}
For the quantum critical part of the susceptibility we use an expression without adjustable parameters:
\begin{eqnarray}\label{QCfit}
\chi_{QC}(H,T)=\left(\frac{\pi-2}{\gamma -1}\right)^{\gamma}\times  \nonumber\\ \left( \left(\frac{T}{\pi}\right)^\gamma \left(1-\left(\frac{T}{\pi}\right)^\gamma \right) + \frac{\displaystyle T^{2\gamma} +H^2-H_c^2 }{\displaystyle 1+(T/\pi)^\gamma } \right)^{-\gamma}
\end{eqnarray}
where the critical transverse field $H_c$ is related to the critical exponent by two conditions:
\begin{equation}\label{constr}
H_c=\frac{\pi -2\gamma}{\gamma -1}, \ \ \ \ \ \frac{\gamma^2+1}{\gamma^2-1}=8-H_c
\end{equation}
or explicitly:
\begin{eqnarray}\label{gamma}
\gamma=\frac{1}{18}\left(\pi-2+\sqrt{328+32\pi+\pi^2} \right) \nonumber\\ H_c=\frac{4+16\pi-2\sqrt{328+32\pi+\pi^2}}{-20+\pi+\sqrt{328+32\pi+\pi^2}}
\end{eqnarray}
Approximately, $H_c=3.03692$ and $\gamma=1.226645$. Recent estimate for $H_c$ is given in ref.\cite{hamer}. Despite being based on poor data in this area, the quantum critical susceptibility eq.(\ref{QCfit}) continues seamlessly onto the line $T=0$. Therefore, $\gamma$ is the critical exponent of the susceptibility of the three dimensional Ising model. The quantum critical susceptibility shows two crossovers as temperature varies at $T_1^*=1/\pi$ and at $T_2^*=\pi$. For the regular part of the susceptibility we try a polynomial. The best fit is given by the expression that depends on temperature sharply above $T^*_2$ and explicitly vanishes at the quantum critical point $(H_c,0)$:
\begin{equation}\label{regular}
\chi_{reg}(H,T)=a\frac{H^2}{H^2_c}\left(\frac{H^2}{H^2_c} -1\right) +b\left(\frac{T}{\pi} \right)^{9\gamma}
\end{equation}
where $a=0.00313$ and $b=0.000115$. This fit promotes the eq.(\ref{QCfit}) as the asymptotic susceptibility at the quantum critical point. Probably, it also indicates a hidden symmetry as though the transverse field Ising Hamiltonian eq.(\ref{Ham}) is the fixed point from the beginning of the renormalization flow. The first term in eq.(\ref{regular}), probably, describes a tail from a crossover at $H^*\sim 7$ where the receding quantum critical susceptibility eq.(\ref{QCfit}) transforms into an one-site susceptibility $\chi(H,T)=1/(2H)$ at the high transverse fields. There is no evidence of the quantum-to-classical crossover in the numerical simulations for graphs with weights $w\leq 11$.

The quantum critical susceptibility eq.(\ref{QCfit}) has a singularity line $(H,T_Q(H))$ on the phase diagram Fig.\ref{fig:diag}. We prove that this line lies above the Ising phase transition line $(H,T_c(H))$. The quantum-to-classical crossover occurs at some line $(H,T^*(H))$, where $\chi_{QC}(T)$ eq.(\ref{QCfit}) transforms into $\chi_{cl}(T)$ eq.(\ref{chiCl}). Approximating $\chi_{QC}(T)$ by a power law like eq.(\ref{chiCl}) and imposing the two continuity conditions, $\chi_{QC}= \chi_{cl}$ and $d\chi_{QC}/dT= d\chi_{cl}/dT$ at $T=T^*(H)$, we find that $T^*-T_c =(7/4\gamma)(T^*-T_Q)$. Since $\gamma<7/4$, we find $T_c(H)<T_Q(H)$.

One objection against the quantum critical susceptibility eq.(\ref{QCfit}) is that usually the location of the critical point on the phase diagram is not universal. In the one-dimensional chain $H_c=1$ due to the duality \cite{fs78}. The change of the sign in eq.(\ref{QCfit}) at small $H$ serves as to separate the quantum from the classical regions whereas it is problematic at high temperature. Also, crossovers in eq.(\ref{QCfit}) differs markedly from the characteristic triangular quantum criticallity region \cite{sachdev}.

\begin{figure}
\includegraphics[width=7cm]{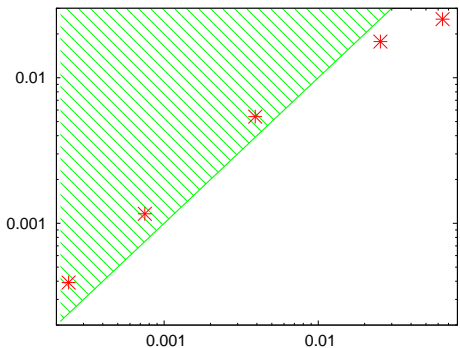}
\caption{\label{fig:fit} Average deviation of the susceptibility data, sorted out into five quality baskets, from the fit vs average variation of the data as the maximum weight increases by one.}
\end{figure}

The following test demonstrates that there is no intrinsic contradiction between our data and the fit eqs.(\ref{chi},\ref{QCfit},\ref{regular}). The quality $\delta(H,T)$ of datum gives us a crude estimate of how far it may potentially vary as the maximum weight increases from $w_{max}=11$ to $w_{max}=\infty$. Averaging all data variations over a basket of points we get a measure for the final potential variation. On the other hand, we can measure the current discrepancy between the data and the fit, at $w_{max}=11$, and average it over the same basket of points. If the current discrepancy is exceeding the potential variation by far it is unlikely that the data and the fit will converge at $w\to\infty$. Alternatively, when the potential variation is exceeding the current discrepancy by far, a special pattern of alternating signs is necessary for the data and the fit to converge that seems also unlikely. The best chance for the data and the fit to converge is when the potential variation approximately equals the current discrepancy. Our data and the fit present widely varying both measures shown in the Fig.\ref{fig:fit}. Remarkably, these two measures are more or less equal, see Fig.\ref{fig:fit}.

In conclusion, using the graph expansion the longitudinal susceptibility of the transverse field Ising model on the square lattice has been calculated numerically. The result has been non-contradictory interpreted in terms of a simple function. Such interpretation might be useful when searching for a corresponding settings in the Einstein general relativity. It is quite possible that our interpretation is erroneous. It is also possible that eq.(\ref{gamma}) gives the correct critical exponent of the susceptibility of the three dimensional Ising model.

I am grateful to SLAC Scientific Computing Services for providing resources for numerical simulations.

\end{document}